\documentclass[11pt,a4paper]{article}
\usepackage[english]{babel}
\usepackage{amsmath,amssymb}
\usepackage{cite}
\usepackage[margin=2.5cm]{geometry}
\usepackage{float}
\usepackage[section]{placeins}

\usepackage{graphicx}
\graphicspath{{figures/}}
\usepackage{mathrsfs}
\usepackage{xcolor}
\usepackage{hyperref}
\hypersetup{colorlinks=true,citecolor=blue,urlcolor=blue,linkcolor=blue}
\makeatletter
\newcommand{\beq}{\begin{equation}}
\newcommand{\eeq}{\end{equation}}
\newcommand{\bea}{\begin{eqnarray}}
\newcommand{\eea}{\end{eqnarray}}

\newcommand{\comm}[1]{}

\usepackage{bm}

\usepackage{booktabs}

\begin{document}
\title{Rotating Black Holes in Perfect Fluid Dark Matter: Horizons, Photon Orbits, and Shadows}
\author{Gamal G. L. Nashed\\[2mm]
\small Centre for Theoretical Physics, The British University in Egypt,\\
\small P.O. Box 43, El Sherouk City, Cairo 11837, Egypt\\
\small Centre for Space Research, North-West University,\\
\small Potchefstroom 2520, South Africa\\
\small \texttt{nashed@bue.edu.eg}}
\date{}
\maketitle

\begin{abstract}
We investigate rotating black holes surrounded by perfect-fluid dark matter (PFDM), described by the parameter \(k\), and determine the Kerr-connected region of the \((a/M,k/M)\) parameter space that contains an event horizon. The critical value \(k_{\mathrm c}(a)\), obtained from the degenerate-horizon conditions \(\Delta(r_{\mathrm c})=\Delta'(r_{\mathrm c})=0\), defines the boundary between black-hole and horizonless configurations. We derive an exact parametric form of this extremal boundary and show that its near-Kerr behavior includes a logarithmic correction; therefore, \(k_{\mathrm c}/M\simeq2.8828(1-a/M)^{1.5377}\) is only an empirical fit over \(0.90\leq a/M\leq0.999\), not an asymptotic scaling law. For horizon-bearing solutions, we calculate unstable spherical photon orbits and the shadow's mean radius, area, and distortion, finding that increasing \(k/M\) shifts the photon region inward and reduces the shadow size. Backward ray tracing through a phenomenological optically thin disk, tested on independent \(41^2\), \(81^2\), and \(161^2\) image grids, confirms this trend: for \(a/M=0.9\) and an inclination of \(17^\circ\), the \(20~\mu\mathrm{as}\) beam-convolved ridge diameter decreases from \(42.47\) to \(37.88~\mu\mathrm{as}\) as \(k/M\) increases from 0 to \(0.08\), with the two finest grids agreeing to within \(0.5\%\). Thus, PFDM produces a consistent contraction from the horizon and photon region to the geometric shadow and image ridge, although these illustrative images do not provide an Event Horizon Telescope constraint because they lack a self-consistent emission model and interferometric likelihood analysis.
\end{abstract}

\section{Introduction}

Dark matter is supported by evidence from galactic rotation curves, gravitational lensing, the cosmic microwave background, and the growth of large-scale structure \cite{Zwicky1933,Rubin:1980zd,Planck:2018vyg}. However, its microscopic nature and distribution near compact objects remain unclear. Because dark matter is detected mainly through its gravitational effects, an important question is whether it can modify the strong-field geometry around black holes and influence photon motion near the event horizon.

Black holes offer an ideal setting for exploring this question. Horizon-scale images of M87\(^{*}\) and Sgr A\(^{*}\), produced by the Event Horizon Telescope (EHT), have made it possible to observe gravity in the strong-field regime \cite{EventHorizonTelescope:2019dse,EventHorizonTelescope:2019ggy,EventHorizonTelescope:2022wkp}. The central brightness depression and surrounding emission are partly determined by null geodesics that travel near unstable spherical photon orbits. Because the resulting critical curve and geometric shadow depend directly on the spacetime metric, black-hole shadows provide a useful way to test general relativity and investigate environmental effects or deviations from the Kerr geometry \cite{Falcke:1999pj,Perlick:2015vta,Cunha:2018gql,Cunha:2018acu}.

{Complementary studies of compact objects in alternative gravitational frameworks have investigated anisotropic stellar configurations in higher-order curvature gravity, mimetic gravity, and $f(R,T)$ gravity~\cite{Nashed:2021gkp,Nashed:2021pkc,Nashed:2023uvk}. These stellar models concern different physical systems from the rotating PFDM black holes considered here and are cited only as broader context for strong-field compact-object research.}

In this context, black holes surrounded by perfect-fluid dark matter (PFDM) provide a phenomenological model in which environmental effects appear as a logarithmic modification of the spacetime metric \cite{Li:2012zx,Xu:2017bpz,Hou:2018avu}. Earlier models of black holes surrounded by anisotropic matter were developed by Kiselev \cite{Kiselev:2002dx}, whereas the geometry considered here is based on the later logarithmic PFDM model. This PFDM contribution modifies both the horizon equation and the null-geodesic potential, and may therefore influence the horizon structure, photon region, gravitational lensing, accretion features, and observable black-hole shadow.

{
The horizon structure of rotating PFDM black holes has
been investigated in earlier studies. In particular,
Rizwan, Jamil, and Jusufi~\cite{Rizwan:2018rgs}
examined the critical spin at fixed PFDM intensity,
identifying the boundary between black-hole and
horizonless configurations. The present analysis
addresses the same extremality boundary from a
complementary perspective: for a prescribed spin,
we determine the corresponding critical PFDM
parameter and obtain an exact parametric
representation of the degenerate-horizon radius,
critical spin, and PFDM intensity. This formulation
also makes it possible to establish the logarithmic
near-Kerr behavior of the extremal boundary and
to distinguish it from the empirical power-law
approximation valid over a finite spin interval.
}

Photon orbits connect the horizon structure to what a distant observer sees. Unstable circular and spherical null trajectories determine the geometric critical curve, incorporating the effects of both black-hole rotation and the surrounding matter distribution \cite{Teo:2003ltt,Perlick:2021aok}. We therefore calculate the radii of prograde and retrograde photon orbits, construct the shadow boundary, and evaluate its mean radius, area, and distortion. All shadow and image calculations are limited to the horizon-bearing region defined by the exact extremal boundary, ensuring that horizonless configurations are not interpreted as black holes.

Recent studies have examined images of rotating PFDM black holes using general-relativistic ray tracing, thin-disk emission models, and comparisons with EHT observations of M87\(^{*}\) and Sgr A\(^{*}\) \cite{Huang:2026xgt}. Polarization properties have also been explored through covariant polarization transport under different magnetic-field configurations \cite{Huang:2026cjn}. While these studies focus on emission physics, polarized intensity, and image-based observables, our work takes a complementary approach. We first establish the exact Kerr-connected extremal boundary and quantify its effects on the photon region and geometric shadow. We then use a simple optically thin ray-tracing model to test whether the predicted shadow contraction also appears in intensity images. Comparisons with EHT angular scales are therefore illustrative only and are not intended as observational fits.

{
Previous studies have also examined the influence
of PFDM on photon trajectories and black-hole
shadow observables. {Xu, Hou, and Wang~\cite{Xu:2017bpz}
constructed and analyzed a rotating PFDM spacetime
with a cosmological constant, providing geometric
background for the present metric.} Hou, Xu, and
Wang~\cite{Hou:2018avu} analyzed the dependence
of rotating black-hole shadows on the PFDM
parameter. Their results demonstrate that the
shadow size and distortion can exhibit different
trends across the parameter space, including
a reduction in shadow size within particular
PFDM regimes. These findings provide an
important reference for the present analysis,
which focuses on horizon-bearing configurations
and quantifies the associated changes in the
prograde and retrograde photon orbits, shadow
radius, area, and distortion. Unlike studies
concentrating primarily on shadow morphology,
the present work explicitly connects these
geometric observables with the exact
Kerr-connected extremality boundary and
supplements the analysis with independently
refined relativistic intensity images.
}

This study makes three main contributions. First, we derive an exact parametric form of the Kerr-connected extremal boundary and its logarithmic behavior near the Kerr limit, clarifying the restricted validity of a simple power-law fit. Second, we show how the allowed PFDM parameter modifies unstable photon orbits and key shadow properties. Third, we connect these geometric predictions to independently ray-traced intensity images and test numerical convergence using multiple screen resolutions and geodesic-invariant diagnostics. This approach traces the effect of PFDM from the event horizon to the image plane while clearly distinguishing geometric predictions from realistic observational constraints.

The paper is organized as follows. Section~\ref{II} examines the horizon structure and derives the exact parametric form of the Kerr-connected extremal boundary, along with a finite-interval empirical fit. Section~\ref{III} presents the unstable photon orbits and their dependence on the PFDM parameter. Section~\ref{IV} constructs the black-hole shadow and evaluates its radius, area, and distortion. Section~\ref{sec:raytracing} shows illustrative relativistic thin-disk intensity images and numerical convergence diagnostics, while Section~\ref{V} compares the predicted geometric angular scale with that reported for M87\(^{*}\). Section~\ref{VI} discusses the physical implications, limitations, and possible extensions, and the main conclusions are summarized in Section~\ref{sec:conclusions}.

\section{Horizon Structure and Extremal PFDM Black Holes}\label{II}

Before examining photon regions and shadow properties, we first establish the geometric structure of the rotating PFDM spacetime \cite{Li:2012zx,Xu:2017bpz}. The existence of an event horizon and the corresponding extremal configurations provide the basis for interpreting the photon and shadow signatures discussed later. This section analyzes the horizon structure, derives the critical PFDM parameter required for horizon existence, and investigates its behavior near extremality.

\subsection{Horizon Structure and Critical PFDM Parameter}

We use geometrized units, \(G=c=1\), throughout. Therefore, the parameters \(M\), \(a\), and \(k\), together with the radial coordinate \(r\), have units of length. In the numerical analysis, all lengths are scaled by the black-hole mass \(M\), after which we set \(M=1\). Thus, the quantities denoted below by \(a\), \(k\), and \(r\) represent the dimensionless ratios \(a/M\), \(k/M\), and \(r/M\), respectively. In Boyer--Lindquist coordinates, the line element for the rotating PFDM spacetime is written as
\begin{align}
ds^2={}&-\left(1-\frac{2m(r)r}{\Sigma}\right)dt^2
-\frac{4am(r)r\sin^2\theta}{\Sigma}\,dt\,d\phi
+\frac{\Sigma}{\Delta}\,dr^2+\Sigma\,d\theta^2 \nonumber\\
&+\left[r^2+a^2+
\frac{2a^2m(r)r\sin^2\theta}{\Sigma}\right]
\sin^2\theta\,d\phi^2,
\label{PFDM_metric}
\end{align}
where
\begin{equation}
\Sigma=r^2+a^2\cos^2\theta,
\qquad
\Delta(r)=r^{2}+a^{2}-2m(r)r,
\qquad
m(r)=M-\frac{k}{2}\ln\!\left(\frac{r}{|k|}\right).
\label{PFDM_functions}
\end{equation}
For $k=0$, the PFDM contribution is defined by its continuous limit,
$k\ln(r/|k|)\to0$, and the Kerr geometry is recovered. The horizon
positions are the positive roots of $\Delta(r)=0$.
Substituting the effective mass function yields
\begin{equation}\label{delta}
\Delta(r)=r^{2}+a^{2}-2Mr
+kr\ln\!\left(\frac{r}{|k|}\right), \quad \mbox{where $r_{\mathrm{ H}}$ satisfies} \quad
\Delta(r_{\mathrm{ H}})=0.
\end{equation}
The horizon equation differs from that of the Kerr spacetime because it includes a logarithmic term proportional to the PFDM parameter \(k\). As a result, both the location and existence of the horizon depend on \(k\). Numerical solutions show that the horizon moves inward as \(k\) increases. For example, when \(a=0.8\), the horizon radius changes from
\begin{equation}
r_{\mathrm{ H}}=1.6000, \quad \mbox{for the Kerr case ($k=0$) to}\quad r_{\mathrm{ H}}=0.9982\quad \mbox{for $k=0.3$.}
\end{equation}
This result confirms that increasing the PFDM parameter shifts the event horizon to a smaller radius. The limiting case occurs when the inner and outer horizons merge into a single degenerate horizon. The corresponding critical PFDM parameter, \(k_{\mathrm c}\), is obtained by solving the simultaneous equations
\begin{equation}
\Delta(r_{\mathrm{ c}})=0, \quad \mbox{and} \quad
\left.\frac{d\Delta}{dr}\right|_{r=r_{\mathrm {c}}}=0.
\end{equation}
For a representative spin $a=0.9$, we obtain
\begin{equation}
k_{\mathrm{ c}}=0.0860938448837392,\quad \mbox{and} \quad
r_{\mathrm{ c}}=0.8579819547489318.
\end{equation}
The horizon becomes degenerate at this point, satisfying
\begin{equation}
\Delta(r_{\mathrm{ c}})=0,
\qquad
\Delta'(r_{\mathrm{ c}})=0.
\end{equation}
Along the Kerr-connected low-$k$ branch, the critical PFDM intensity
decreases rapidly as the spin approaches the extremal Kerr limit:
\begin{align}
&k_{\mathrm{ c}}(0.90)=0.08609,\quad
k_{\mathrm{ c}}(0.95)=0.02861,\nonumber\\
&k_{\mathrm{ c}}(0.98)=0.00841,\quad
k_{\mathrm{ c}}(0.99)=0.00355.
\end{align}
\begin{figure}[H]
\centering
\includegraphics[width=0.62\textwidth]
{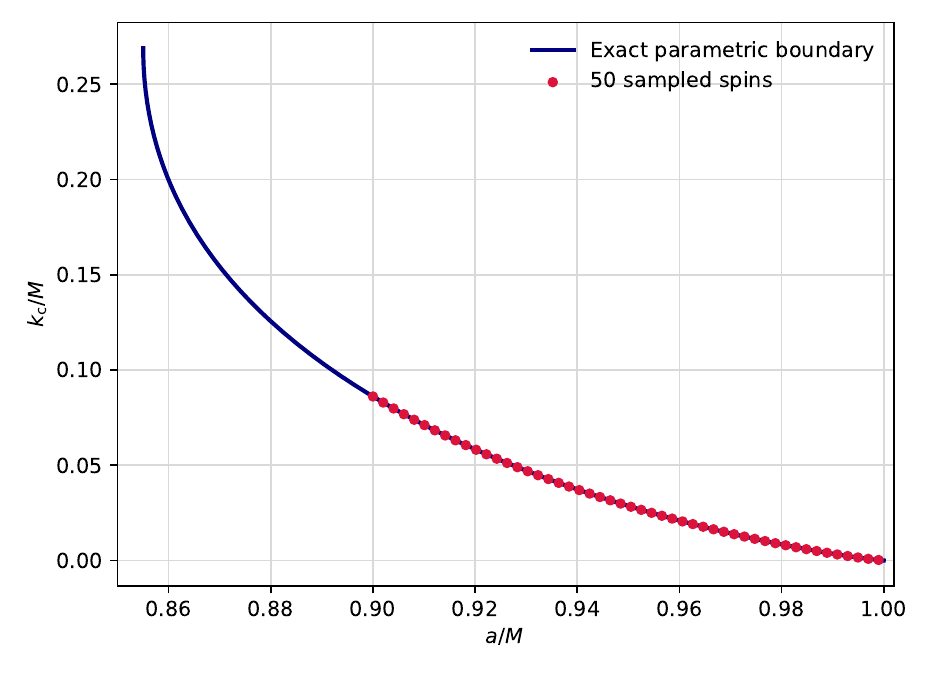}
\caption{Exact Kerr-connected extremal boundary in the
$(a/M,k_{\mathrm{c}}/M)$ plane. The solid curve follows from the
parametric representation in Eq.~\eqref{exact_parametric_boundary}.
The red points show the 50 uniformly sampled spin values over
$0.90\leq a/M\leq0.999$ used in the finite-interval fit. On the
displayed low-$k$ branch, black-hole configurations satisfy
$k<k_{\mathrm{c}}(a)$.}
\label{fig:exact_extremal_boundary}
\end{figure}
For each spin value \(a\), the extremal limit is determined by solving \(\Delta(r_{\mathrm c})=0\) and \(\Delta'(r_{\mathrm c})=0\) simultaneously. These equations yield the critical PFDM parameter \(k_{\mathrm c}(a)\) and the radius \(r_{\mathrm c}\) at which the inner and outer horizons merge. We also verified every parameter set used in the numerical analysis by solving \(\Delta(r_{\mathrm H})=0\) and selecting the largest positive root. For example, \((a,k)=(0.8,0.3)\) gives \(r_{\mathrm H}\simeq0.9982\), confirming that this configuration possesses an event horizon and represents a rotating PFDM black hole.

\subsection{Exact Parametric Form of the Extremal Boundary}

The extremal curve can be written exactly in parametric form. From
Eq.~\eqref{delta}, the second extremality condition is
\begin{equation}
2r_c-2M+k_c\left[\ln\!\left(\frac{r_c}{k_c}\right)+1\right]=0,
\label{extremal_derivative}
\end{equation}
where this subsection considers the positive-$k$ branch. Multiplying
Eq.~\eqref{extremal_derivative} by $r_c$ and using
$\Delta(r_c)=0$ eliminates the logarithm and gives
\begin{equation}
a_c^2=r_c(r_c+k_c).
\label{exact_extremal_identity}
\end{equation}
Introducing the positive parameter
\begin{equation}
x=\frac{k_c}{r_c},
\label{extremal_parameter_x}
\end{equation}
the derivative condition can be solved directly. The extremal boundary
is therefore represented by
\begin{align}
\frac{r_c(x)}{M}
&=\frac{2}{2+x(1-\ln x)}, \nonumber\\
\frac{k_c(x)}{M}
&=\frac{2x}{2+x(1-\ln x)}, \nonumber\\
\frac{a_c(x)}{M}
&=\frac{2\sqrt{1+x}}{2+x(1-\ln x)}.
\label{exact_parametric_boundary}
\end{align}
Equations~\eqref{exact_extremal_identity} and
\eqref{exact_parametric_boundary} are exact and reproduce the numerical
solutions of the simultaneous extremality conditions. The branch
connected continuously to extremal Kerr corresponds to
$x\to0^+$ and, up to the minimum of $a_c$, to
$0<x\leq1/\mathrm{e}$. At its endpoint,
\begin{equation}
x=\frac{1}{\mathrm{e}},\qquad
\frac{k_c}{M}=\frac{1}{1+\mathrm{e}},\qquad
\frac{a_c}{M}=\sqrt{\frac{\mathrm{e}}{1+\mathrm{e}}}\simeq0.855.
\label{extremal_turning_point}
\end{equation}
This branch structure is consistent with Ref.~\cite{Rizwan:2018rgs}, where the same boundary was described in terms of a critical spin for a fixed PFDM parameter. In contrast, our formulation provides \(r_{\mathrm c}\), \(k_{\mathrm c}\), and \(a_{\mathrm c}\) simultaneously through the single parameter \(x\), while isolating the near-extremal branch relevant to the shadow calculations. The exact representation also determines the leading behavior close to the extremal Kerr limit. Expanding Eq.~\eqref{exact_parametric_boundary} as \(x\to0^+\) gives
\begin{equation}
1-\frac{a_c}{M}
=-\frac{x\ln x}{2}+O(x),
\qquad
\frac{k_c}{M}=x+O\!\left(x^2|\ln x|\right).
\label{exact_near_kerr_asymptotics}
\end{equation}
Hence the true leading approach contains a logarithmic correction,
$1-a_c/M\sim-(k_c/2M)\ln(k_c/M)$, rather than an exact power law.

\subsection{Finite-Interval Fit of the Extremal Boundary}

To characterize the extremal boundary over the finite interval
$0.90\leq a\leq0.999$, we fit the numerical values of $k_c(a)$
using the empirical power-law form
\begin{equation}
k_c(a)=A(1-a)^p.
\label{empirical_kc_fit}
\end{equation}
Using 50 uniformly sampled spin values, we obtain
\begin{equation}
A=2.8828\pm0.1029,
\qquad
p=1.5377\pm0.0140,
\label{empirical_fit_parameters}
\end{equation}
and hence
\begin{equation}
k_c(a)\simeq2.8828(1-a)^{1.5377}.
\label{empirical_kc_result}
\end{equation}
Within the fitted interval, this expression provides only a numerical summary of the extremal boundary and should not be interpreted as an exact asymptotic scaling law. For each sampled value of \(a\), we obtained the critical PFDM parameter by solving the simultaneous extremality conditions \(\Delta(r_{\mathrm c})=0\) and \(\Delta'(r_{\mathrm c})=0\). Over \(0.90\leq a\leq0.999\), the empirical power-law fit gives \(R^2=0.99842\), indicating excellent agreement within this finite range. Although the fitted exponent, \(p=1.5377\pm0.0140\), is close to \(3/2\), Eq.~\eqref{exact_near_kerr_asymptotics} shows that the exact near-Kerr behavior includes a logarithmic term and is therefore not a pure \(3/2\) power law.
\begin{figure}[H]
\centering
  \includegraphics[width=0.62\textwidth]{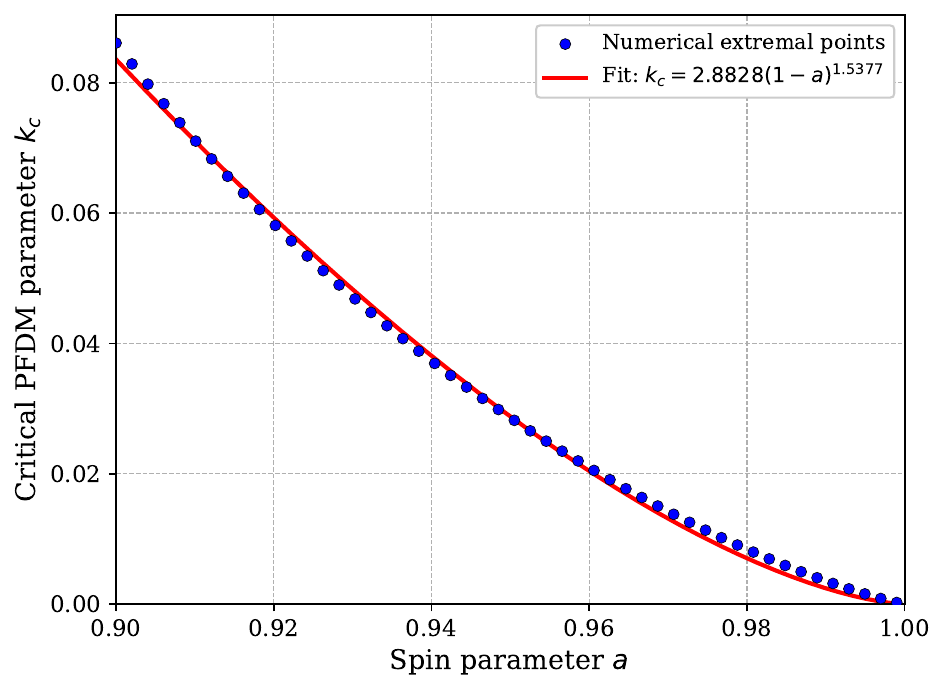}
\caption{Critical PFDM parameter $k_c$ as a function of the spin
parameter $a$ over the interval $0.90\leq a\leq0.999$. The points
represent numerical solutions of the extremality conditions
$\Delta(r_c)=0$ and $\Delta'(r_c)=0$, while the solid curve shows
the empirical finite-interval fit
$k_c(a)=2.8828(1-a)^{1.5377}$.}
\label{fig:2}
\end{figure}
The numerical solutions show that $k_c$ becomes very small toward the
upper end of the sampled spin interval. The fit is useful for numerical
interpolation there, while the exact asymptotic behavior is supplied by
Eq.~\eqref{exact_near_kerr_asymptotics}.

\section{Photon Region in Rotating PFDM Spacetimes}\label{III}

The unstable circular photon orbits determine the boundary of the
black-hole shadow and play a central role in shaping black-hole
imaging observables. We solve the photon-orbit equation numerically for several choices of
$a$ and $k$, considering both prograde and retrograde equatorial
orbits. A solution is retained as physical only when it satisfies
\begin{equation}
r_{\mathrm{ ph}}>r_{\mathrm{ H}}.
\end{equation}
In addition to this condition, the circular photon orbits relevant for the
shadow boundary must be unstable. Therefore, after determining the photon
orbit radii from the circularity conditions, we also verify the instability
condition
\begin{equation}
\left.\frac{d^2\mathcal{R}(r)}{dr^2}\right|_{r=r_{\mathrm {ph}}}>0,
\end{equation}
where $\mathcal{R}(r)$ is the radial photon potential. This ensures that the selected
photon orbits correspond to the unstable light rings that form the boundary of
the black-hole shadow. For the parameter values considered below, both the
prograde and retrograde photon-orbit branches lie outside the event horizon
and satisfy the above instability condition.

The numerical results show that increasing $k$ moves both photon orbits
toward the black hole. For example, when $a=0.8$, the radius of the
prograde orbit changes from
\begin{align}
&r_{\mathrm{ ph}}^{+}=1.8111, \quad \mbox{for $k=0$ to} \quad r_{\mathrm{ ph}}^{+}=1.0580, \nonumber\\
&\mbox{for $k=0.3$, corresponding to a reduction of approximately $41.6\%$.}
\end{align}

Similarly, the retrograde photon orbit decreases from
\begin{align}
r_{\mathrm{ ph}}^{-}=3.8188, \quad \mbox{to} \quad
r_{\mathrm{ ph}}^{-}=2.9563,\nonumber\\ \mbox{corresponding to a reduction of approximately $22.6\%$.}
\end{align}

The change is larger for the prograde orbit than for the retrograde
one. Hence, the two sides of the photon region do not respond equally
to $k$. This unequal displacement changes the asymmetry of the photon
region and is then reflected in the appearance of the black-hole
shadow.

For a more direct comparison, we introduce the relative change in each
photon-orbit radius as
\begin{equation}
\Delta r_{\mathrm{ ph}}^{\pm}
=
\frac{
r_{\mathrm{ ph}}^{\pm}(k)-r_{\mathrm{ ph}}^{\pm}(0)
}{
r_{\mathrm{ ph}}^{\pm}(0)
}
\times100\%.
\end{equation}
\begin{figure}[H]
\centering
\includegraphics[width=0.72\textwidth]
{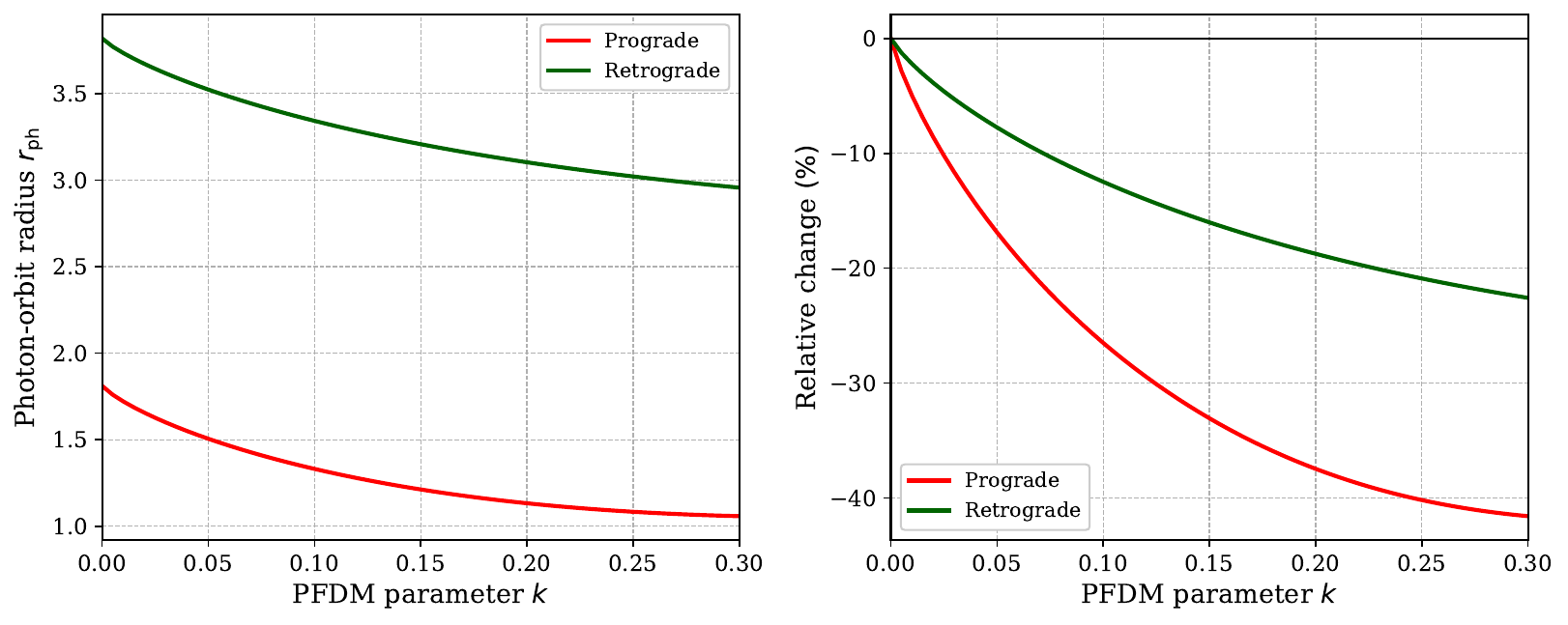}
\caption{Equatorial photon-orbit radii for a rotating PFDM black
hole with $a/M=0.8$. The left panel shows the prograde and
retrograde photon-orbit radii as functions of the PFDM parameter
$k/M$, while the right panel presents their relative changes with
respect to the Kerr limit $k=0$. Both radii decrease over the
displayed interval, with the prograde branch showing the stronger
response. All displayed photon orbits lie outside the corresponding
outer event horizon.}
\label{fig:photon_orbits_a08}
\end{figure}
This difference between the two branches is related to the combined
action of rotation and PFDM on photon motion. The prograde orbit lies
closer to the horizon because the photons travel in the same direction
as the black-hole spin. In this inner region, its position changes more
strongly when $k$ is varied. The prograde radius therefore decreases by
a larger fraction than the retrograde radius.
\section{Shadow Structure and Observational Signatures}\label{IV}

The black-hole shadow represents one of the most direct observational
manifestations of the underlying spacetime geometry. Its boundary is
determined by unstable spherical photon orbits, which separate photon
trajectories captured by the black hole from those that escape to a
distant observer. Consequently, any modification of the spacetime
metric is expected to leave a measurable imprint on the shadow
morphology.

In the rotating PFDM spacetime, the logarithmic correction introduced
by the dark matter parameter $k$ alters the photon region and modifies
the critical null geodesics that generate the shadow boundary.
Therefore, the influence of PFDM propagates naturally from the horizon
structure to the photon region and ultimately to the observable shadow.
A quantitative analysis of the shadow can thus provide valuable
information about the gravitational effects of dark matter in the
strong-field regime.

In this section, we construct the shadow boundary using the celestial
coordinates of unstable spherical photon orbits and investigate the
resulting shadow observables. Particular attention is devoted to the
shadow radius, area, and distortion parameter, which characterize the
size and shape of the shadow and provide potentially observable
signatures of PFDM. By examining the dependence of these quantities on
the PFDM parameter $k$, we identify the dominant geometric effects of
dark matter on black-hole imaging observables.
\subsection{Shadow Boundary and Celestial Coordinates}

The shadow boundary is generated by unstable spherical photon orbits satisfying
\begin{equation}
\mathcal{R}(r)=0,
\qquad
\frac{d\mathcal{R}(r)}{dr}=0,
\end{equation}
where the radial potential is
\begin{equation}
\mathcal{R}(r)
=
\left[(r^{2}+a^{2})E-aL_{z}\right]^{2}
-
\Delta(r)
\left[
\mathcal{Q}
+
\left(L_{z}-aE\right)^{2}
\right].
\end{equation}
Introducing the impact parameters
\[
\xi=\frac{L_z}{E},
\qquad
\eta=\frac{\mathcal{Q}}{E^2},
\]
where $E$ denotes the conserved photon energy, $L_z$ is the conserved
axial component of the photon angular momentum, and $\mathcal{Q}$ is
the Carter constant associated with the hidden symmetry of the
spacetime.
The critical impact parameters become
\begin{equation}
\xi(r)=
\frac{
(r^2+a^2)\Delta'(r)-4r\Delta(r)
}{
a\,\Delta'(r)
},\qquad \mbox{and} \qquad
\eta(r)=
\frac{
16r^2\Delta(r)
}{
[\Delta'(r)]^2
}
-
(\xi-a)^2.
\end{equation}
For an observer located at inclination angle $\theta_0$, the celestial coordinates of the shadow boundary are
\begin{equation}\label{alphabeta}
\alpha=
-\frac{\xi}{\sin\theta_0}, \quad \mbox{and} \quad
\beta=
\pm
\sqrt{
\eta
+a^2\cos^2\theta_0
-\xi^2\cot^2\theta_0
}.
\end{equation}

These relations provide the direct connection between the PFDM-modified spacetime geometry and observable shadow morphology.
The shadow boundary was obtained by numerically sampling the unstable
spherical photon orbits and evaluating Eq.~\eqref{alphabeta}. The resulting
celestial coordinates $(\alpha,\beta)$ were then used to construct a
discrete representation of the shadow contour, from which the shadow
radius, area, and distortion observables were computed.

\subsection{Shadow Radius and Area}

\begin{figure}[H]
\centering
\includegraphics[width=0.62\textwidth]{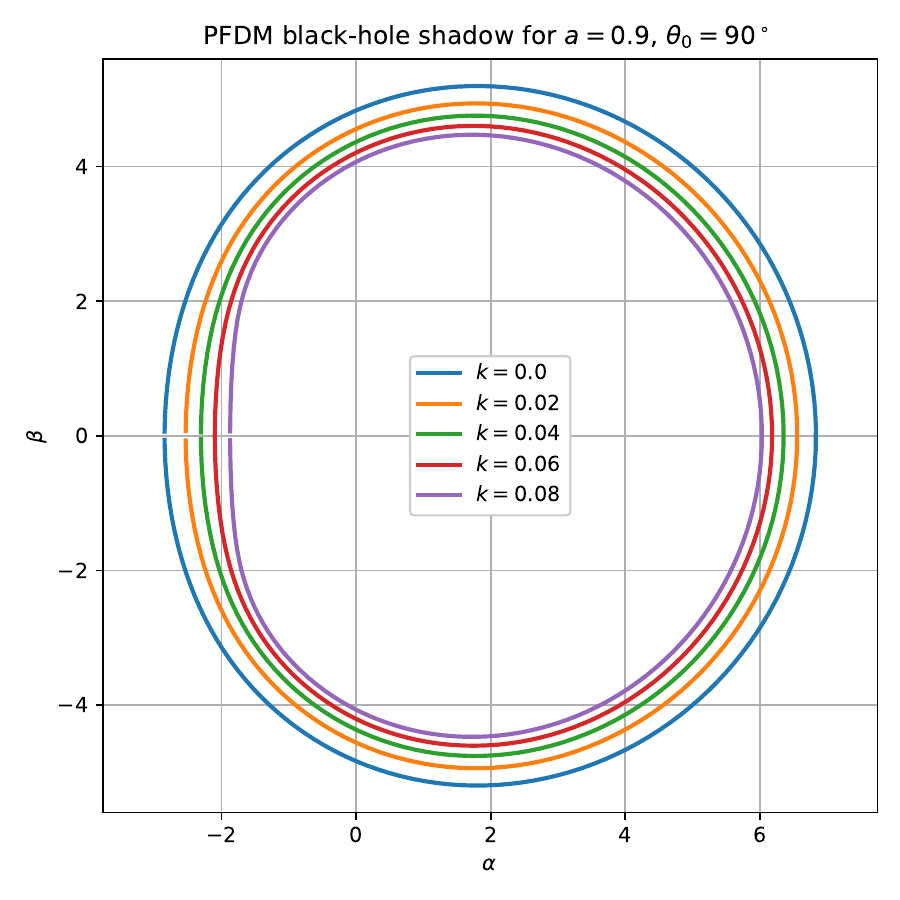}
\caption{Shadow contours of a rotating PFDM black hole for
$a/M=0.9$ and $\theta_0=90^\circ$. The spin-induced asymmetry
is clearly visible, while increasing $k/M$ systematically reduces
the shadow size. All displayed values satisfy
$k<k_{\mathrm c}(0.9)\simeq0.08609$ and therefore correspond to
black-hole configurations possessing an event horizon.}
\label{fig:shadow_contours}
\end{figure}

\begin{figure}[H]
\centering
\includegraphics[width=0.62\textwidth]{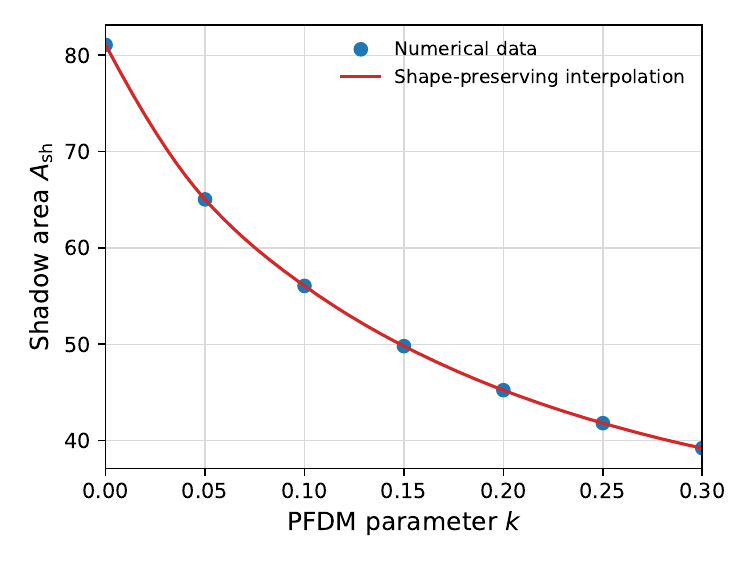}
\caption{Shadow area $A_{\mathrm {sh}}$ as a function of the PFDM
parameter $k$ for $a=0.8$ and $\theta_0=90^\circ$. The event
horizon exists throughout the displayed interval
$0\leq k\leq0.3$.}
\label{fig:Ash_vs_k}
\end{figure}

We consider two representative configurations. First, the
near-extremal case $a=0.9$ and $\theta_0=90^\circ$ is examined
over the admissible interval $0\leq k\leq0.08$, which lies below
the critical value $k_c(0.9)\simeq0.08609$. This configuration is
used to display the shadow contours in
Fig.~\ref{fig:shadow_contours} and to construct
Table~\ref{tab:shadow_observables}. Second, the configuration
$a=0.8$ and $\theta_0=90^\circ$ is used to investigate the wider
interval $0\leq k\leq0.3$. For this configuration, the existence
of an outer horizon has been verified throughout the complete
parameter interval. Its shadow area and relative radius variation
are displayed in Figs.~\ref{fig:Ash_vs_k} and
\ref{fig:shadrad}, respectively.

The mean shadow radius is defined as
\begin{equation}
R_{\mathrm {sh}}
=
\frac{1}{N}
\sum_{i=1}^{N}
\sqrt{(\alpha_i-\alpha_c)^2+(\beta_i-\beta_c)^2},
\label{mean_shadow_radius}
\end{equation}
where $(\alpha_c,\beta_c)$ denotes the geometric center of the shadow.  In the numerical evaluation, the spherical-orbit radius is sampled
uniformly between the prograde and retrograde equatorial light-ring
radii. Consequently, the quantity in Eq.~\eqref{mean_shadow_radius} should be
understood as a sampling-dependent mean radius.

 For the equatorial observer considered here, the shadow is
symmetric with respect to the horizontal axis, and its center is
defined by
\begin{equation}
\alpha_c=\frac{\alpha_{\mathrm {max}}+\alpha_{\mathrm {min}}}{2},
\qquad
\beta_c=0,
\label{shadow_center}
\end{equation}
where $\alpha_{\mathrm {max}}$ and $\alpha_{\mathrm {min}}$ are the rightmost
and leftmost celestial coordinates of the shadow boundary,
respectively. In Eq.~\eqref{mean_shadow_radius}, $N$ denotes the
number of ordered numerical points used to represent the closed
shadow contour.

For the wider-range configuration $a=0.8$ and
$\theta_0=90^\circ$, the mean shadow radius decreases from
\begin{equation}
R_{\mathrm {sh}}(k=0)=5.1282,
\qquad
R_{\mathrm {sh}}(k=0.3)=3.5912,
\label{shadow_radius_values}
\end{equation}
corresponding to a reduction of approximately $30.0\%$.
For the same configuration, the shadow area is computed from
\begin{equation}
A_{\mathrm {sh}}
=
\frac{1}{2}
{\bigl|}
\sum_{i=1}^{N}
\left(
\alpha_i\beta_{i+1}
-
\beta_i\alpha_{i+1}
\right)
{\bigr|},
\label{shadow_area}
\end{equation}
where the contour points are ordered consecutively and the cyclic
identification
$(\alpha_{N+1},\beta_{N+1})=(\alpha_1,\beta_1)$ is understood.
Numerically, we obtain
\begin{equation}
A_{\mathrm {sh}}(k=0)=81.08,
\qquad
A_{\mathrm {sh}}(k=0.3)=39.22.
\label{shadow_area_values}
\end{equation}
Combining the horizon, photon-region, and shadow analyses, the overall chain of PFDM effects can be summarized as
\begin{equation}
k\uparrow
\;\Longrightarrow\;
r_{\mathrm{ H}}\downarrow
\;\Longrightarrow\;
r_{\mathrm{ ph}}\downarrow
\;\Longrightarrow\;
R_{\mathrm{ sh}}\downarrow.
\end{equation}
Therefore, over the parameter intervals examined here, increasing
$k$ shifts the outer horizon and photon region toward smaller
radii and reduces the apparent black-hole shadow.

\subsection{Shadow Distortion}
To further quantify the deformation of the shadow, we also compute the shadow distortion parameter $\delta_s$. Following the standard Hioki--Maeda construction~\cite{Hioki:2009na}, one defines a reference circle passing through the top, bottom, and rightmost points of the shadow. If $(\alpha_t,\beta_t)$ denotes the upper point, $\alpha_r$ the rightmost point, and $\alpha_p$ the leftmost point of the shadow boundary, the reference radius is
\begin{equation}
R_s=\frac{(\alpha_t-\alpha_r)^2+\beta_t^2}{2|\alpha_r-\alpha_t|}.
\end{equation}
The leftmost point of the reference circle is then
\begin{equation}
\tilde{\alpha}_p=\alpha_r-2R_s,
\end{equation}
and the distortion parameter is defined as
\begin{equation}
\delta_s=\frac{\alpha_p-\tilde{\alpha}_p}{R_s}.
\end{equation}
This quantity measures the deviation of the shadow from a circular shape and therefore provides a complementary observable to the shadow radius and area.

\begin{table}[H]
\centering
\caption{Shadow observables for a rotating PFDM black hole with
$a=0.9$ and $\theta_0=90^\circ$. All the reported values satisfy
$k<k_c(0.9)\simeq0.08609$ and therefore correspond to black-hole configurations possessing an event horizon.}
\begin{tabular}{cccc}
\hline
$k$ & $R_{\mathrm{sh}}$ & $A_{\mathrm{sh}}$ & $\delta_s$ \\
\hline
0.00 & 5.1087 & 79.59 & 0.13868 \\
0.02 & 4.8491 & 71.24 & 0.16132 \\
0.04 & 4.6675 & 65.60 & 0.18118 \\
0.06 & 4.5181 & 61.01 & 0.20317 \\
0.08 & 4.3942 & 57.01 & 0.23494 \\
\hline
\end{tabular}
\label{tab:shadow_observables}
\end{table}
Table~\ref{tab:shadow_observables} shows that increasing $k$ reduces both the mean shadow
radius $R_{\mathrm {sh}}$ and the shadow area $A_{\mathrm {sh}}$, while also
increasing the distortion parameter $\delta_s$. For the
configuration $a=0.9$ and $\theta_0=90^\circ$, the mean shadow
radius decreases from $R_{\mathrm {sh}}=5.1087$ at $k=0$ to
$R_{\mathrm {sh}}=4.3942$ at $k=0.08$, corresponding to a reduction of
approximately $14.0\%$. Over the same interval, the shadow area
decreases from $A_{\mathrm {sh}}=79.59$ to $A_{\mathrm {sh}}=57.01$, a
reduction of approximately $28.4\%$.

The distortion parameter increases from $\delta_s=0.13868$ at
$k=0$ to $\delta_s=0.23494$ at $k=0.08$. Although its absolute
change is approximately $0.096$, this corresponds to a relative
increase of about $69.4\%$. Therefore, PFDM affects both the size
and the shape of the shadow in this near-extremal configuration.
In absolute terms, the dominant change is the contraction of the
shadow radius and area, while the distortion exhibits a
substantial relative increase.

\subsection{Impact of PFDM on Shadow Observables}

The results presented in the previous sections demonstrate that the
PFDM parameter $k$ produces a systematic
modification of the observable shadow characteristics of a rotating
black hole. Since the shadow boundary is determined by unstable
spherical photon orbits, any PFDM-induced change in the spacetime
geometry is directly reflected in the observable shadow properties.

The mathematical origin of these effects can be traced directly to the
PFDM modification of the metric function,
\begin{equation}
\Delta(r;k)=r^2+a^2-2Mr+k r\ln\left(\frac{r}{|k|}\right).
\end{equation}
The dependence of $\Delta(r;k)$ on $k$ enters the critical impact
parameters through
\begin{align}
&\xi(r;k)=
\frac{(r^2+a^2)\Delta'(r;k)-4r\Delta(r;k)}
{a\Delta'(r;k)},\quad \mbox{and} \nonumber\\
&\eta(r;k)=
\frac{16r^2\Delta(r;k)}{[\Delta'(r;k)]^2}
-\left[\xi(r;k)-a\right]^2 .
\end{align}
Therefore, changing $k$ modifies the critical photon trajectories
that generate the shadow boundary.

For an observer at inclination angle $\theta_0$, these impact
parameters determine the celestial coordinates
\begin{align}
\alpha(r;k)
&=
-\frac{\xi(r;k)}{\sin\theta_0},
\nonumber\\
\beta(r;k)
&=
\pm\left[
\eta(r;k)
+a^2\cos^2\theta_0
-\xi^2(r;k)\cot^2\theta_0
\right]^{1/2}.
\label{shadow_coordinates_k}
\end{align}
Thus, the PFDM parameter modifies the apparent shadow contour through
the direct dependence of $\alpha$ and $\beta$ on $\Delta(r;k)$.

The shadow observables are then computed from the resulting discrete
contour points $(\alpha_i,\beta_i)$. The mean shadow radius is
\begin{equation}
R_{\mathrm {sh}}(k)=
\frac{1}{N}\sum_{i=1}^{N}
\sqrt{[\alpha_i(k)-\alpha_c(k)]^2+
[\beta_i(k)-\beta_c(k)]^2},
\end{equation}
while the shadow area is obtained from
\begin{equation}
A_{\mathrm {sh}}(k)=
\frac{1}{2}
{\bigl|}
\sum_{i=1}^{N}
\left[
\alpha_i(k)\beta_{i+1}(k)
-\beta_i(k)\alpha_{i+1}(k)
\right]
{\bigr|}.
\end{equation}
The distortion parameter is similarly expressed as
\begin{equation}
\delta_s(k)=
\frac{\alpha_p(k)-\tilde{\alpha}_p(k)}
{R_s(k)} .
\end{equation}
These relations show explicitly that all shadow observables inherit
their $k$-dependence from the PFDM-modified metric function
$\Delta(r;k)$.

The influence of PFDM first appears at the level of the horizon
structure. As demonstrated by the horizon equation in
Sec.~\ref{II}, increasing $k$ reduces the outer-horizon radius within
each of the parameter intervals examined here. The extremality
conditions determine the Kerr-connected critical value $k_c(a)$ that
separates the low-$k$ black-hole branch from an adjacent horizonless
interval. The reduction of the horizon radius
indicates that PFDM alters the effective gravitational environment
experienced by null geodesics.

These modifications propagate to the photon region through the
conditions governing unstable spherical photon orbits. The numerical
analysis shows that both the prograde and retrograde photon-orbit
radii decrease as the PFDM intensity increases. The contraction of
the photon region implies that the critical photon trajectories
responsible for forming the shadow boundary occur at smaller radial
distances from the black hole.

As a consequence, the observable shadow size decreases systematically
with increasing PFDM parameter. For the representative configurations
considered in this work, both the shadow radius $R_{\mathrm {sh}}$ and the
shadow area $A_{\mathrm {sh}}$ exhibit a monotonic reduction as $k$
increases. This behavior indicates that the PFDM correction shifts the
critical photon region toward smaller radial coordinates over the
parameter interval considered. The resulting contraction of the
shadow is therefore one clear geometric signature of PFDM in the
present model.

The distortion parameter $\delta_s$ also changes appreciably with
$k$. For the near-extremal configuration reported in
Table~\ref{tab:shadow_observables}, its absolute change is about
$0.096$, corresponding to a relative increase of approximately
$69.4\%$. Thus, PFDM affects both the overall shadow size and its
asymmetry, although these effects are expressed through different
observables.

Combining the horizon, photon-region, and shadow analyses, the
overall sequence of PFDM effects can be summarized schematically as

\begin{equation}
k \uparrow
\Longrightarrow
r_{\mathrm {H}} \downarrow
\Longrightarrow
r_{\mathrm {ph}} \downarrow
\Longrightarrow
R_{\mathrm {sh}} \downarrow
\Longrightarrow
A_{\mathrm {sh}} \downarrow .
\end{equation}

\section{Illustrative relativistic intensity images}
\label{sec:raytracing}

The numerical configurations used in this work have distinct purposes.
The equatorial $a/M=0.8$ sequence extends to $k/M=0.3$ and illustrates
the wider geometric trend, whereas the equatorial $a/M=0.9$ sequence is
restricted to $k/M\leq0.08$ by the Kerr-connected extremal boundary and
quantifies the near-extremal shadow. The present image calculation uses
$a/M=0.9$ and the M87$^{*}$-motivated inclination $17^\circ$. Results
from these configurations should therefore be compared as complementary
tests, not as points on one common parameter sequence.

The geometric shadow does not by itself specify the observed brightness
distribution. To test whether the contraction found above is retained
in a simple emission model, we backward trace null geodesics from a
distant observer through the rotating PFDM metric. The calculation uses
Hamilton's equations with conserved photon energy normalized to
$E=1$ and axial angular momentum $\xi=L_z/E$. Each equatorial crossing
of an optically thin disk contributes
\begin{equation}
 \Delta I_{\mathrm {obs}}=w_n g^3 j(r),\qquad
 g=\frac{1}{u^t(1-\Omega\xi)},\qquad
 w_n=0.65^n ,
\label{eq:image_intensity}
\end{equation}
where $\Omega$ and $u^t$ are those of a prograde equatorial circular
geodesic and $n$ labels the crossing order. We retain at most three
crossings. The adopted phenomenological emissivity,
\begin{equation}
 j(r)\propto \frac{1}{r^2}
 \exp\!\left[-\frac{\ln^2(r/5M)}{2(0.55)^2}\right],
\label{eq:image_emissivity}
\end{equation}
extends from the outer prograde marginally stable orbit to $30M$.
Equations~\eqref{eq:image_intensity} and \eqref{eq:image_emissivity}
define a phenomenological specific-intensity proxy with the standard $g^3$ weighting.
They do not constitute a frequency-integrated radiative-transfer
calculation.

{
The adopted accretion-disk prescription is motivated by the need
for a simple, physically plausible emission distribution that
allows the influence of the PFDM-modified spacetime on the
observed image to be examined independently of detailed plasma
microphysics. The emissivity profile in Eq.~(\ref{eq:image_emissivity}) combines an
inverse-square radial dependence with a log-normal factor,
providing a smooth, centrally concentrated emission region
and suppressing contributions from large radii.
{The characteristic radial scale $5M$ sets the reference
radius of the log-normal factor, while the logarithmic
width $0.55$ controls the spread of emission over disk
radii. Together with the $r^{-2}$ factor, these choices
produce a smooth emissivity concentrated toward the
strong-field region. Both numerical values are illustrative
phenomenological choices rather than parameters inferred
from observations or derived from a self-consistent
accretion-flow model. The maximum of the unrestricted
profile occurs at $r=5M\exp[-2(0.55)^2]\simeq 2.73M$;
if this lies inside the adopted disk inner edge, the
maximum over the emitting disk is instead at that edge.}
 The disk
extends from the outer prograde marginally stable orbit
to $30M$, while the redshift factor $g^3$ in Eq.~(\ref{eq:image_intensity})
accounts for the transformation of specific intensity
along photon trajectories. This simplified prescription
enables a controlled comparison of the image morphology
for different PFDM parameters using the same emission
model. Consequently, the resulting images are intended
to illustrate the effects of spacetime geometry rather
than provide a detailed astrophysical description of
the accretion plasma.}
\subsection{Numerical implementation and validation}

Rays are launched backward from an observer at $r_{\mathrm{obs}}=100M$
on a square screen of width $24M$. We integrate Hamilton's equations
with relative and absolute tolerances $2\times10^{-7}$ and
$2\times10^{-9}$, respectively. A trajectory is terminated when it
crosses a numerical capture surface immediately outside the outer
horizon, escapes the computational domain, reaches the maximum affine
parameter $500M$, or supplies the retained three disk crossings. The
disk inner edge is the outer prograde marginally stable orbit and its
outer edge is $30M$. The same stopping rules and emission prescription
are used at every screen resolution.

{
The observer is placed at $r_{\mathrm{obs}}=100M$,
sufficiently far from the near-horizon region to provide
a convenient numerical setup while keeping the ray-tracing
integration computationally manageable. This choice allows
the photon trajectories to probe the strong-field region
responsible for the main relativistic features of the
observed images. Nevertheless, placing the observer at a
finite distance may introduce corrections to the mapping
between local photon directions and the corresponding
asymptotic image-plane coordinates, potentially affecting
the apparent image size and brightness distribution.
These effects may also depend on the PFDM parameter
through the radial behavior of the spacetime metric.
To ensure a consistent comparison, the same observer
position and image-screen prescription are employed
for all PFDM configurations. Although this procedure
provides a common basis for identifying the influence
of PFDM on the image observables, a systematic
investigation at larger values of $r_{\mathrm{obs}}$
would be required to quantify the remaining
finite-distance corrections.
{In particular, fixing $r_{\mathrm{obs}}$ across
models does not by itself cancel finite-distance effects,
since the PFDM correction to the large-radius metric
contains a term proportional to
$(k/r)\ln(r/|k|)$. The reported image observables
should therefore be interpreted for the stated finite
observer location rather than as demonstrated
asymptotic-observer values.}
}

Two invariant diagnostics provide checks independent of the image
observables. On representative rays from the finest endpoint
calculations, the largest absolute null-Hamiltonian residual is
$4.83\times10^{-5}$ for Kerr and $5.18\times10^{-8}$ for $k/M=0.08$;
the corresponding largest relative Carter-constant variations are
$4.26\times10^{-7}$ and $8.34\times10^{-8}$. Table~\ref{tab:geodesic_validation}
summarizes these checks.
{The absolute null-Hamiltonian residuals differ substantially
between the two endpoint samples, whereas the relative variations
of the Carter constant remain below $5\times10^{-7}$ in both cases.
Because the Hamiltonian residual is reported in absolute rather than
normalized form, the two maxima cannot by themselves be interpreted
as directly comparable relative integration errors. Moreover, these
diagnostics cover five representative rays per endpoint, rather than
the full image screens. The independent screen-resolution refinements
reported below assess the stability of the extracted image observables,
but do not replace an integration-tolerance convergence test. The
origin of the difference between the absolute Hamiltonian residuals
has not been independently established.}

\begin{table}[H]
\caption{Geodesic-invariant checks for five representative rays in each
$161^2$ endpoint calculation.}
\label{tab:geodesic_validation}
\centering
\begin{tabular}{ccc}
\toprule
$k/M$ & $\max|\mathcal H|$ & $\max|\Delta\mathcal Q/\mathcal Q|$\\
\midrule
0.00 & $4.83\times10^{-5}$ & $4.26\times10^{-7}$\\
0.08 & $5.18\times10^{-8}$ & $8.34\times10^{-8}$\\
\bottomrule
\end{tabular}
\end{table}

\begin{figure}[H]
\centering
\includegraphics[width=0.88\textwidth]{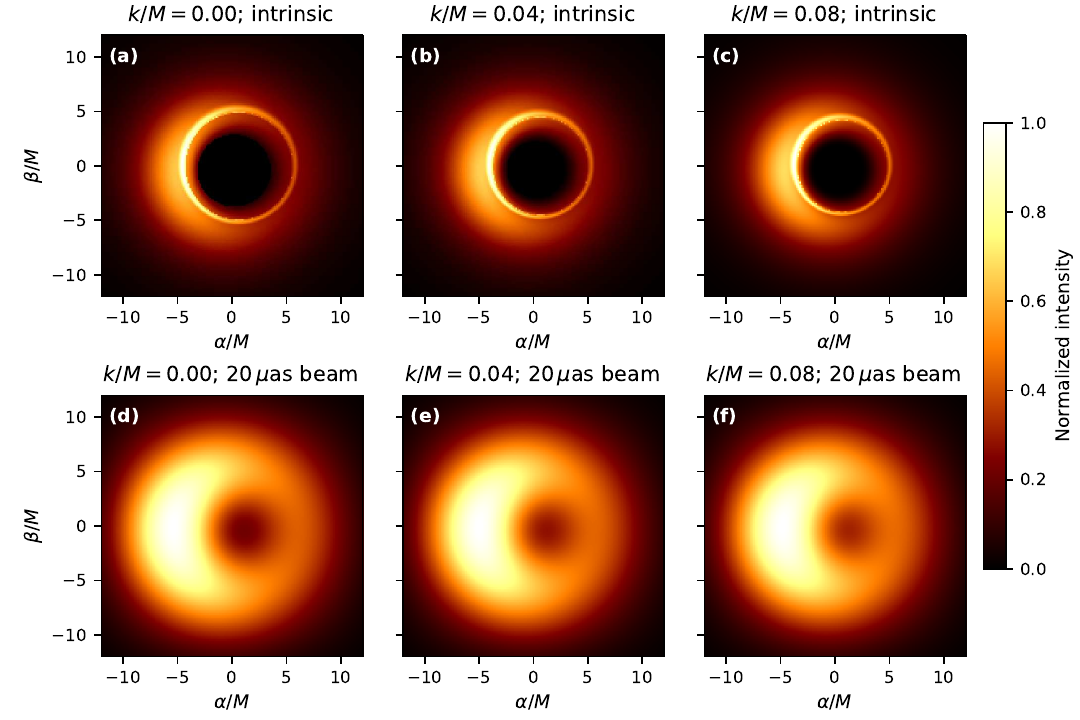}
\caption{Illustrative PFDM thin-disk images for $a/M=0.9$,
inclination $17^\circ$, and screen resolution $161^2$. Columns show
$k/M=0$, $0.04$, and $0.08$; the upper and lower rows show the intrinsic
and $20~\mu\mathrm{as}$ beam-convolved images, respectively. All panels
use normalized intensity. Increasing $k$ contracts the bright ridge,
while the finite beam broadens the emission and partially fills the
central depression.}
\label{fig:raytracing_comparison}
\end{figure}

We use $a/M=0.9$, inclination $17^\circ$, an observer radius $100M$,
a $24M$ field of view, and a $161\times161$ screen. The five values
$k/M=0$, $0.02$, $0.04$, $0.06$, and $0.08$ all lie below
$k_{\mathrm c}(0.9)\simeq0.08609$ and therefore possess an event
horizon. To
illustrate finite angular resolution, each normalized image is also
convolved with a circular Gaussian of full width at half maximum
$20~\mu{\mathrm {as}}$, using $\theta_g=GM/(Dc^2)=3.8~\mu{\mathrm {as}}$.
The ring radius is measured from the peak intensity along 360 radial
screen directions. Table~\ref{tab:raytracing_diagnostics} gives the
resulting ridge diameters and approaching-to-receding brightness ratios.
The brightness-asymmetry ratio is defined by
\begin{equation}
\mathcal A
=
\frac{
\left\langle I_{\mathrm{ridge}}\right\rangle_{\mathrm{app}}
}{
\left\langle I_{\mathrm{ridge}}\right\rangle_{\mathrm{rec}}
},
\end{equation}
where the averages are evaluated along the approaching and receding
halves of the image ridge, respectively. With the screen orientation
used in Fig.~\ref{fig:raytracing_comparison}, the approaching and
receding averages are evaluated over the $\alpha<0$ and $\alpha>0$
halves of the ridge, respectively.
\begin{table}[H]
\caption{Diagnostics of the illustrative intensity images. Diameters
are measured from the radial intensity ridge. ``Beam'' denotes
convolution with a $20~\mu{\mathrm { as}}$ circular Gaussian.}
\label{tab:raytracing_diagnostics}
\centering
\begin{tabular}{ccccc}
\toprule
$k/M$ & $r_{\mathrm {H}}/M$ & $d_{\mathrm {int}}$ ($\mu$as)
& $d_{\mathrm { beam}}$ ($\mu$as) & $\mathcal A_{\mathrm {beam}}$\\
\midrule
0.00 & 1.436 & 38.73 & 42.47 & 1.696\\
0.02 & 1.288 & 36.85 & 40.60 & 1.681\\
0.04 & 1.176 & 35.32 & 39.41 & 1.665\\
0.06 & 1.070 & 34.30 & 38.56 & 1.653\\
0.08 & 0.947 & 33.28 & 37.88 & 1.641\\
\bottomrule
\end{tabular}
\end{table}

\begin{figure}[H]
\centering
\includegraphics[width=0.72\textwidth]{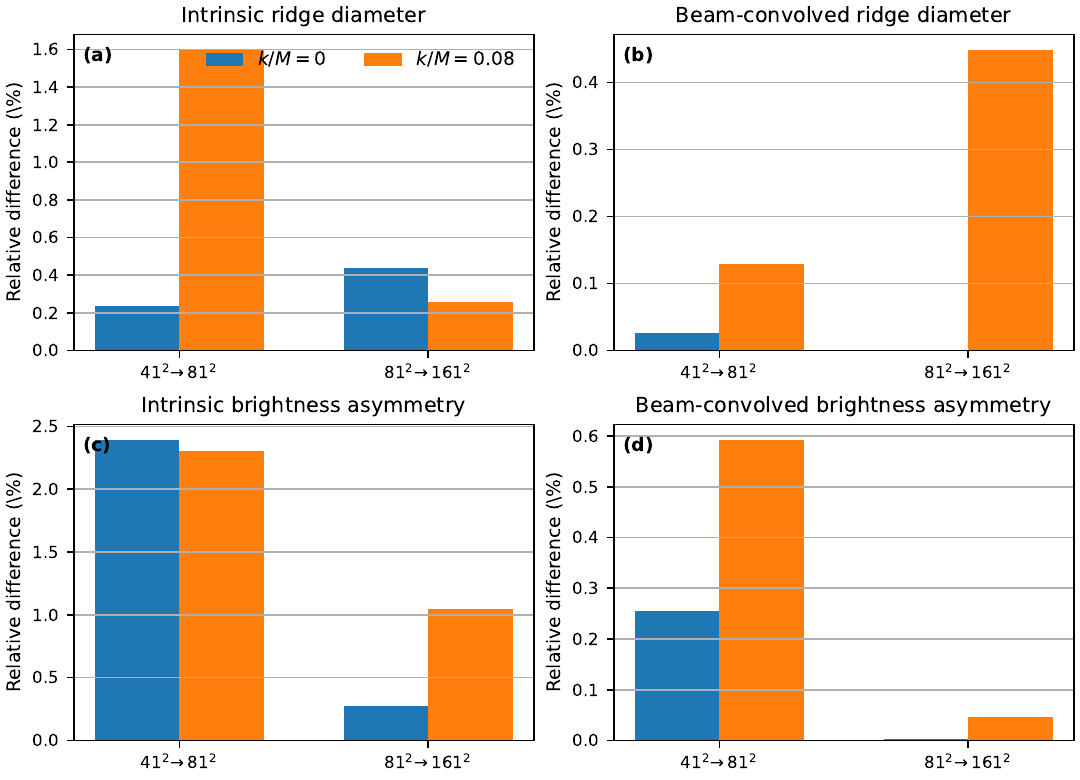}
\caption{Fractional changes under the independent refinements
$41^2\rightarrow81^2$ and $81^2\rightarrow161^2$ for the Kerr
reference and the largest PFDM value in the image sequence. Panels show
the intrinsic and beam-convolved ridge diameters and brightness
asymmetries. Every grid uses newly traced rays rather than a subsample
of a finer image.}
\label{fig:raytracing_diagnostics}
\end{figure}

\begin{table}[H]
\caption{Percentage changes under independent screen refinement.
Absolute percentage differences are measured relative to the finer
grid.}
\label{tab:image_convergence}
\centering
\begin{tabular}{cccccc}
\toprule
$k/M$ & refinement & $\Delta d_{\rm int}$ & $\Delta d_{\rm beam}$
& $\Delta\mathcal A_{\rm int}$ & $\Delta\mathcal A_{\rm beam}$\\
 & & (\%) & (\%) & (\%) & (\%)\\
\midrule
0.00 & $41^2\!\to81^2$  & 0.234 & 0.026 & 2.395 & 0.255\\
0.00 & $81^2\!\to161^2$ & 0.439 & 0.000 & 0.273 & 0.003\\
0.08 & $41^2\!\to81^2$  & 1.599 & 0.129 & 2.305 & 0.593\\
0.08 & $81^2\!\to161^2$ & 0.256 & 0.449 & 1.043 & 0.046\\
\bottomrule
\end{tabular}
\end{table}

The intrinsic ridge diameter falls by approximately $14.1\%$ between
$k=0$ and $k=0.08$, while the beam-convolved value falls by
approximately $10.8\%$. The beam-convolved brightness ratio decreases
from $1.696$ to $1.641$.

We tested the two endpoint models by tracing independent $41^2$,
$81^2$, and $161^2$ screens. For $k=0$, the beam-convolved diameters
are $42.480$, $42.469$, and $42.469~\mu{\mathrm {as}}$, respectively,
whereas the intrinsic values are $38.988$, $38.896$, and
$38.726~\mu{\mathrm {as}}$. For $k=0.08$, the corresponding
beam-convolved values are $37.657$, $37.706$, and
$37.876~\mu{\mathrm {as}}$, and the intrinsic values are $32.834$,
$33.367$, and $33.282~\mu{\mathrm {as}}$. Thus, between the $81^2$ and
$161^2$ grids, all four endpoint diameter diagnostics change by less
than $0.5\%$. The beam-convolved brightness ratios change by less than
$0.05\%$ over the same refinement, and the intrinsic ratios change by
at most about $1.1\%$. The extracted intrinsic ring width is more
resolution-sensitive and is therefore not used as a precision
observable. The brightness ratio itself also depends directly on the
assumed emissivity and circular-flow prescription.

This exercise closes part of the gap between a geometric critical curve
and an intensity image, but it does not supply an observational
constraint. In particular, the model omits absorption, synchrotron
microphysics, magnetic fields, time variability, scattering, and the
interferometric sampling and likelihood used by the EHT. The images
should consequently be viewed as a controlled demonstration that the
PFDM-induced size trend survives one simple relativistic emission and
beam model.

\section{Illustrative Comparison with the
\texorpdfstring{M87$^{*}$}{M87*} Observation}\label{V}

The geometric shadow size predicted by the PFDM spacetime can be
compared illustratively with the angular scale measured by the
Event Horizon Telescope. For a black hole of mass $M$ at distance
$D$, the theoretical angular diameter associated with the
geometric shadow is
\begin{equation}
d_{\mathrm {sh}}^{\mathrm {th}}(k)
=
2R_{\mathrm {sh}}(k)\theta_g,
\qquad
\theta_g=\frac{GM}{Dc^2},
\label{theoretical_angular_shadow}
\end{equation}
where $R_{\mathrm {sh}}(k)$ is the dimensionless mean shadow radius.

The EHT measurement, however, refers to the diameter of the
observed emission ring rather than to a direct, model-independent
measurement of the geometric shadow boundary. Relating the
measured ring diameter to $d_{\mathrm {sh}}^{\mathrm {th}}$ requires an image
model that accounts for the accretion flow, radiative transfer,
instrumental response, and the displacement between the emission
ring and the critical curve. Therefore, the comparison presented
below is intended only to estimate the magnitude of the
PFDM-induced change in angular size; it is not used to derive a
definitive confidence interval for $k$. A more complete treatment was
recently carried out in Ref.~\cite{Huang:2026xgt}, which used
general-relativistic ray tracing of thin accretion disks at 87 and
230~GHz, constructed synthetic intensity images, and confronted the
model with EHT measurements of both M87$^{*}$ and Sgr~A$^{*}$. Our
geometric comparison is deliberately more limited and should be read as
complementary to that radiative-transfer analysis, not as an equivalent
observational constraint.

For an illustrative numerical comparison, we use the angular
scales reported by the EHT Collaboration for M87$^{*}$~\cite{EventHorizonTelescope:2019dse,EventHorizonTelescope:2019ggy},
\begin{equation}
d_{\mathrm {ring}}^{\mathrm {obs}}=42\pm3~\mu{\mathrm {as}},
\qquad
\theta_g=3.8\pm0.4~\mu{\mathrm {as}}.
\label{M87_angular_scales}
\end{equation}
Here $d_{\mathrm {ring}}^{\mathrm {obs}}$ denotes the measured emission-ring
diameter. It should not be identified directly with the geometric
shadow diameter without an appropriate emission-model
calibration.

Using the shadow radii in Table~\ref{tab:shadow_observables} and the central value
$\theta_g=3.8~\mu{\mathrm {as}}$, the corresponding theoretical angular
diameters are
\begin{align}
d_{\mathrm {sh}}^{\mathrm {th}}(0)
&=38.83~\mu\mathrm {as},
&
d_{\mathrm {sh}}^{\mathrm {th}}(0.02)
&=36.85~\mu\mathrm {as},
\nonumber\\
d_{\mathrm {sh}}^{\mathrm {th}}(0.04)
&=35.47~\mu\mathrm {as},
&
d_{\mathrm {sh}}^{\mathrm {th}}(0.06)
&=34.34~\mu\mathrm {as},
\nonumber\\
d_{\mathrm {sh}}^{\mathrm {th}}(0.08)
&=33.40~\mu\mathrm {as}.
\label{illustrative_shadow_diameters}
\end{align}
These values show that, for the fixed configuration \(a=0.9\) and \(\theta_0=90^\circ\), the theoretical geometric shadow diameter decreases systematically as \(k\) increases. They indicate the possible size of the PFDM effect but do not provide a statistical constraint on \(k\). We therefore do not construct a likelihood from the quoted values of \(d_{\mathrm{ring}}^{\mathrm{obs}}\) and \(\theta_g\), since they are not independent, model-free measurements of the geometric shadow. In particular, the EHT estimate of \(\theta_g\) already relies on a calibration between the observed emission ring and the underlying spacetime scale. Treating their uncertainties as independent Gaussian errors would ignore this covariance and could produce an artificially strong constraint on \(k\).

The values in Eq.~\eqref{illustrative_shadow_diameters} should be regarded only as an illustrative comparison for \(a=0.9\) and \(\theta_0=90^\circ\). They show that increasing \(k\) decreases the predicted geometric shadow diameter over the allowed interval \(0\leq k\leq0.08\). A statistically meaningful constraint on \(k\) would require a joint analysis of the PFDM parameter, spin, inclination, mass, distance, accretion-flow model, and instrumental response within an appropriate EHT likelihood framework. Since larger values of \(k\) systematically reduce the shadow radius and area, they predict smaller apparent shadow diameters; precision shadow measurements may therefore help constrain PFDM effects near black holes. However, such an inference must account for uncertainties in black-hole properties, viewing geometry, and emission physics. In this work, \(k\) is treated as a phenomenological parameter, with emphasis placed on its effects on the horizon, photon region, and shadow observables. Overall, PFDM produces systematic changes in both the size and shape of the shadow, including a substantial relative increase in distortion for the near-extremal configuration considered here. Future high-resolution black-hole imaging may therefore provide a useful probe of dark-matter effects near compact objects. To quantify this impact, we define the relative variation in the shadow radius as

\begin{equation}
\Delta R_{\mathrm{sh}}(k)
=
\frac{R_{\mathrm{sh}}(k)-R_{\mathrm{sh}}(0)}
     {R_{\mathrm{sh}}(0)}
\times100\%.
\end{equation}

\begin{figure}[H]
\centering
\includegraphics[width=0.62\textwidth]{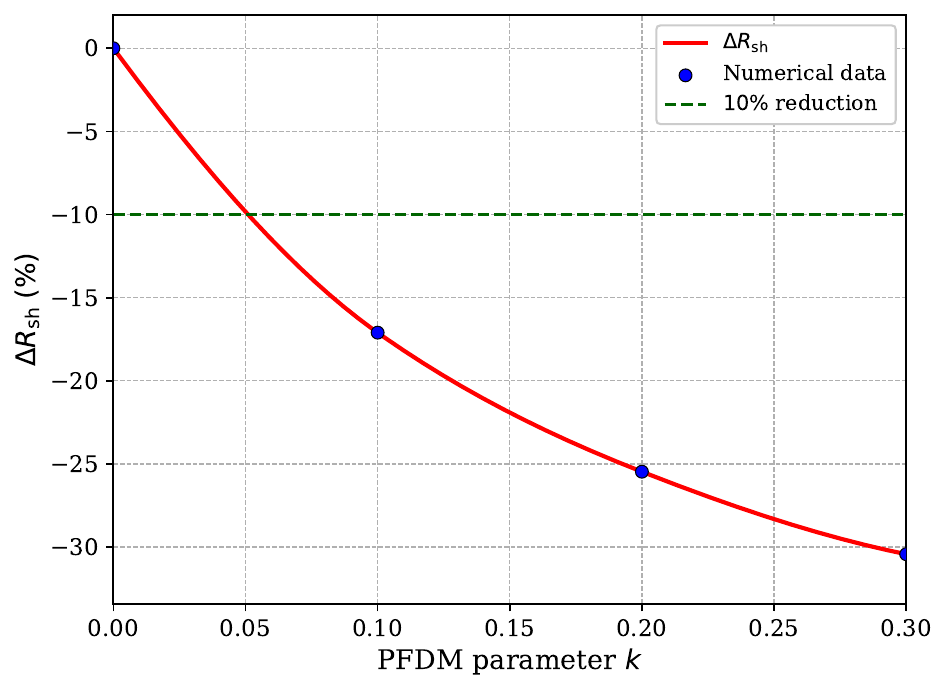}
\caption{Relative variation of the mean shadow radius,
$\Delta R_{\mathrm {sh}}(k)=
[R_{\mathrm {sh}}(k)-R_{\mathrm {sh}}(0)]/R_{\mathrm {sh}}(0)\times100\%$,
for $a=0.8$ and $\theta_0=90^\circ$. The blue points represent
the numerical results, while the red curve is a
shape-preserving interpolation included as a guide to the eye.
The mean shadow radius decreases over the displayed interval,
reaching a reduction of approximately $30.0\%$ at $k=0.3$.
The dashed line indicates an illustrative $10\%$ reduction.}
\label{fig:shadrad}
\end{figure}

This quantity gives the percentage change in shadow size relative to the Kerr case, \(k=0\). Because the shadow radius decreases monotonically as the PFDM intensity increases, \(\Delta R_{\mathrm{sh}}\) directly indicates the observational precision needed to detect PFDM effects in horizon-scale images. Table~\ref{tab:shadow_observables} and Fig.~\ref{fig:shadrad} show this response for two representative cases. For the near-extremal configuration \(a=0.9\) and \(\theta_0=90^\circ\), the mean shadow radius decreases by about \(5.1\%\), \(8.6\%\), \(11.6\%\), and \(14.0\%\) at \(k=0.02\), \(0.04\), \(0.06\), and \(0.08\), respectively. Since all of these values satisfy \(k<k_c(0.9)\simeq0.08609\), they correspond to black-hole configurations with event horizons. Figure~\ref{fig:shadrad} shows a broader parameter range for \(a=0.8\) and \(\theta_0=90^\circ\), where an event horizon exists throughout \(0\leq k\leq0.3\) and the mean shadow radius decreases by about \(30.0\%\) at \(k=0.3\). The distortion parameter, reported separately for the near-extremal case in Table~\ref{tab:shadow_observables}, also shows a substantial relative change.
\section{Discussion}\label{VI}

The analysis provides a unified description of how PFDM affects the strong-field geometry from the event horizon to the observed image. The logarithmic term in Eq.~\eqref{delta} modifies both the horizon equation and the null-geodesic potential. As a result, it changes the horizon location, shifts the unstable photon orbits, and alters the size and shape of the black-hole shadow. For the configurations considered here, increasing \(k\) systematically reduces the characteristic radial scales, from the horizon and photon region to the shadow and image ridge.

Our first main result is the exact parametric representation of the Kerr-connected extremal boundary in Eq.~\eqref{exact_parametric_boundary}. This boundary is equivalent to the critical-spin relation obtained in Ref.~\cite{Rizwan:2018rgs}, but is expressed here in the inverse form needed to determine the largest allowed PFDM parameter for a given spin. For a fixed near-extremal spin, crossing the low-\(k\) branch of this boundary removes the event horizon and leads to a neighboring horizonless region. This distinction is important because the complete positive-\(k\) extremal curve contains the turning point given in Eq.~\eqref{extremal_turning_point}. Along the Kerr-connected branch, \(k_{\mathrm c}\) approaches zero as \(a\) approaches the extremal Kerr limit. Equation~\eqref{exact_near_kerr_asymptotics} further shows that this limit contains a logarithmic correction. Therefore, although Eq.~\eqref{empirical_kc_result} accurately fits the numerical results over \(0.90\leq a\leq0.999\), it should be interpreted as a finite-range interpolation rather than a true asymptotic power law.

The photon region is directly affected by the same PFDM modification. As \(k\) increases, both prograde and retrograde spherical photon-orbit radii move inward, with a stronger contraction on the prograde side. Because the prograde orbit lies closer to the event horizon, it is more sensitive to the combined effects of frame dragging and the PFDM contribution. This behavior agrees with earlier studies of PFDM black holes \cite{Xu:2017bpz,Hou:2018avu} and confirms that changes in the horizon geometry are accompanied by corresponding changes in null-geodesic motion.

{ A more detailed comparison with previous shadow studies
helps clarify the range of validity of these results.
Hou, Xu, and Wang~\cite{Hou:2018avu} found that
the shadow radius and distortion exhibit different
behaviors depending on the PFDM intensity. In
particular, below a characteristic turning point,
increasing the PFDM parameter reduces the shadow
size while enhancing its distortion, whereas the
opposite trends can occur beyond that point.
Our results are consistent with the former behavior
over the horizon-bearing parameter intervals
considered here. For $a/M=0.9$, the mean shadow
radius decreases from $5.1087M$ to $4.3942M$
as $k/M$ increases from $0$ to $0.08$, while
the distortion parameter increases from $0.13868$
to $0.23494$. These results quantify the PFDM
effect within the selected near-extremal regime
and should not be extrapolated to the entire
parameter space. Furthermore, the present
analysis connects the changes in shadow
observables to the exact extremality boundary
and the corresponding inward displacement
of the unstable photon orbits.}

These orbital shifts lead directly to observable changes in the geometric shadow. Within the horizon-bearing parameter range, the mean shadow radius and area decrease monotonically as \(k\) increases, while the shadow shape becomes more distorted. For \(a=0.9\) and an equatorial observer, the distortion parameter increases from \(\delta_s=0.13868\) in the Kerr case to \(\delta_s=0.23494\) at \(k=0.08\). This corresponds to an absolute increase of approximately \(0.096\), or a relative increase of about \(69.4\%\). Thus, the shadow radius and area directly quantify the PFDM-induced contraction, whereas the distortion parameter may show a particularly strong fractional response for rapidly rotating black holes.

The illustrative ray-tracing results show that the geometric contraction remains visible under the simplified emission model used here. For \(a/M=0.9\) and \(\theta_0=17^\circ\), the \(20~\mu\mathrm{as}\) beam-convolved ridge diameter decreases from \(42.47~\mu\mathrm{as}\) at \(k/M=0\) to \(37.88~\mu\mathrm{as}\) at \(k/M=0.08\). The endpoint diameters obtained with the \(81^2\) and \(161^2\) image screens differ by less than \(0.5\%\), supporting numerical convergence of this trend. However, this agreement does not establish that the emission model is physically complete.

{
These image calculations complement the recent
investigation by Huang et al.~\cite{Huang:2026xgt},
who studied rotating PFDM black-hole images
using relativistic ray tracing and thin-disk
emission models, including comparisons with
EHT observations. While that investigation
emphasized the observational appearance of
the emitting region, the present analysis
concentrates on the relation between the
horizon-existence boundary, photon-region
structure, geometric shadow, and image
morphology. An additional aspect of our
numerical treatment is the use of independent
image-grid refinements to assess the stability
of the extracted ridge diameters and
brightness-asymmetry measures. These
comparisons establish the complementary
scope of the two approaches without
identifying the present illustrative images
with fully calibrated observational predictions.
}

The comparison with M87\(^{*}\) must therefore be interpreted cautiously. The EHT emission-ring diameter is not identical to the geometric shadow diameter, since their relationship depends on the accretion flow, radiative transfer, observing response, and source parameters. Although the present images include gravitational redshift, Doppler effects, multiple disk crossings, and Gaussian beam convolution, they rely on a prescribed optically thin emissivity. They do not include absorption, polarized radiative transfer, magnetic-field structure, or time-dependent plasma dynamics \cite{Perlick:2015vta}. In addition, uncertainties in mass, distance, spin, inclination, and instrumental response are not jointly explored. The analysis therefore demonstrates the direction and approximate scale of the PFDM effect, but it does not provide a statistically valid upper bound on \(k\).

A direct observational test will require combining the exact horizon-bearing parameter domain derived here with a self-consistent radiative-transfer model and an interferometric likelihood analysis. This approach could constrain both the source properties and the PFDM contribution using observations of M87\(^{*}\) and Sgr A\(^{*}\). Further extensions could include higher-order terms in the near-Kerr asymptotic expansion, polarized emission, and photon-ring substructure. Such developments would help determine which signatures of PFDM-induced geometric contraction remain observable after astrophysical and instrumental uncertainties are taken into account.

\section{Conclusions}\label{sec:conclusions}
Taken together, the results show a consistent trend {within the horizon-bearing parameter intervals examined here}:
\[
k\uparrow
\quad\Longrightarrow\quad
r_{\mathrm{H}}\downarrow,\qquad
r_{\mathrm{ph}}\downarrow,\qquad
R_{\mathrm{sh}}\downarrow,\qquad
A_{\mathrm{sh}}\downarrow .
\]
{For the horizon-bearing parameter intervals investigated in this study, increasing the PFDM parameter \(k\) decreases the event-horizon radius, the prograde and retrograde photon-orbit radii, and the geometric shadow radius and area.} The overall shadow size exhibits the clearest absolute response, while the distortion may show a large relative change. Future black-hole imaging with improved angular resolution may therefore help probe possible PFDM effects near the event horizon. The main analytic result is the exact parametric Kerr-connected extremal boundary and its logarithmic near-Kerr behavior. The main numerical result is that the predicted geometric contraction remains visible in a simplified relativistic thin-disk image and is stable under independent screen-resolution refinement. Since the analysis does not include a self-consistent plasma model or an interferometric likelihood, these findings identify a robust model trend rather than an observational constraint on \(k\).

\section{Data Availability}

The numerical scripts and derived data needed to reproduce the
ray-traced images, convergence diagnostics, and tables are available
from the corresponding author upon reasonable request.



\end{document}